\documentclass[11pt]{article}

\pdfoutput=1
\usepackage[bottom]{footmisc}
\usepackage{jheppub}

\usepackage{amsfonts}
\usepackage{graphicx}
\usepackage{amsmath}
\usepackage{amssymb}

\usepackage[normalem]{ulem}
\usepackage[utf8]{inputenc}

\usepackage[english]{babel}
\usepackage[autostyle]{csquotes}
\MakeOuterQuote{"}
\usepackage{indentfirst}
\usepackage[dvipsnames]{xcolor}

\title{On the viability of minimal Ho\v{r}ava gravity} 
\author{Ted Jacobson and Pranav Pulakkat}
\affiliation{Maryland Center for Fundamental Physics,\\University of Maryland, College Park, MD 20742, USA}
\emailAdd{jacobson@umd.edu}
\emailAdd{pranavp@umd.edu}
\date{}
\newcommand{\dpar}{\delta_{\parallel}}

\newcommand{\h}{\mathcal{H}}

\def\a{\alpha}

\def\t{\tau}
\def\l{\lambda}

\def\mmHg{$\mathrm{m^{2}Hg}$}
\def\mHg{$\mathrm{mHg}$}

\usepackage{hyperref}

\begin{document}

\flushbottom
\abstract{
Ho\v{r}ava gravity is a Lorentz-violating modification of general relativity (GR) with a preferred spacelike foliation.
Observational evidence has put strong constraints on the parameter values in this model, so that the remaining viable sector is well-characterized by the Newton constant and a single additional parameter. We analyze this restricted theory, which is called minimal Ho\v{r}ava gravity ($\mathrm{mHg}$), from the Hamiltonian point of view. 
We find that in order to 
eliminate a pathological mode that
is unstable at high frequencies and
strongly coupled at low frequencies
the theory must be further restricted so that the slices of the preferred foliation each have constant mean curvature. 
We dub this theory \textit{minimal minimal Ho\v{r}ava gravity} ($\mathrm{m^2Hg}$).
It has two regimes; one in which the mean curvature is time independent, in which case it is equivalent to a particular foliation of GR with nonzero cosmological constant, and another in which 
the mean curvature is time dependent. In the latter there is an infinite propagation
speed since the lapse evolves via 
an elliptic equation,
and the theory thus 
differs from GR in a peculiar
nonlocal fashion.
 To   probe the viability of the sector with time dependent mean curvature, we study in detail the problem of spherical collapse of a thin dust shell. The solution is  determined unambiguously from initial conditions
until the slice
on which the shell first contracts to a point. Afterwards, there is an instantaneously propagating spherical mode with undetermined evolution, demonstrating a failure of the Cauchy problem.}

\maketitle

\section{Introduction}

Ho\v{r}ava(-Lifshitz) gravity \cite{Horava:2009uw}
is a modification of general relativity in which a preferred spacelike foliation of spacetime is introduced and foliation-preserving diffeomorphism symmetry is imposed. This symmetry structure allows for higher spatial derivative terms in the action, and thus a potentially better behaved UV behaviour of the quantum theory.  
In the IR limit, where more than two derivatives are neglected, 
the theory is characterized, apart from the gravitational constant, 
by three parameters, $c_a, c_\sigma, c_\theta$, which are the coupling constants for the terms in the action quadratic in the acceleration,
shear, and expansion of the timelike congruence normal to the 
preferred foliation. Matter is assumed to be insensitive to the preferred foliation, so that it couples only to the spacetime metric. 
The parameter $c_\sigma$ governs the deviation 
of the speed of gravitational waves from the 
speed of electromagnetic waves, and the observation that both gravitational
and electromagnetic waves generated by the coalescence of two neutron stars
arrived at Earth within within about one second of each other
places an extremely strong bound, $|c_\sigma|\lesssim 10^{-15}$.
Parametrized post-Newtonian and other observational constraints then
place the limit $0<c_a\lesssim 10^{-5}$~\cite{EmirGumrukcuoglu:2017cfa}.
It thus becomes interesting to explore the limiting case, where only $c_\theta$ is nonzero, which has been 
referred to as minimal Ho\v{r}ava Gravity (\mHg) \cite{minimal}. 

It should be acknowledged at the outset, however, that 
we are here looking at a theory that is 
finely tuned. Both the assumption that matter is insensitive
to the preferred foliation (i.e., does not couple to the aether, and hence has locally Lorentz invariant dynamics), and that the couplings $c_\sigma$ and 
$c_a$ are zero (or extremely small compared to 1)
are unstable to quantum loop effects, since 
there is no symmetry protecting these tunings. 
For the Lorentz symmetry, several potential protection mechanisms have been studied \cite{Pospelov:2010mp,Bednik:2013nxa,Kharuk:2015wga,Baggioli:2024vza,Marakulin:2016net,Marakulin:2019yxs}.
For the restriction on the couplings, we are not
aware of any protection mechanism.
In the present paper, we set these issues aside,
and focus only on the classical viability of the
theory.

Existing work has shown that, for asymptotically flat spacetimes, the equations of \mHg\;reduce to those of vacuum GR expressed with respect to a foliation for which 
the mean extrinsic curvature $K$ vanishes on all slices \cite{consistency}. 
The present universe, however, is not asymptotically flat. This raises the question
whether \mHg\;in a cosmological setting might be a viable theory exhibiting interesting
effects not included in general relativity. One motivation for pursuing this 
is the outstanding puzzles of cosmology, but we are also interested simply in 
exploring this peculiar feature in the landscape of possible modifications of general relativity.

In this study we
establish that, for general boundary conditions, $K$ need not be constant 
on slices of the preferred foliation. 
However, such solutions exhibit 
pathological (unstable, and strongly coupled) dynamics, due to the presence
of a ``half-mode''. This finding is 
consistent with previous studies 
for the case where only the coupling parameter $c_a$ vanishes \cite{Li:2009bg, Blas:2009yd}
We therefore take the viewpoint that
the theory is viable only 
when $K$ is restricted to be constant
on each slice of the preferred foliation
but depends on time. 
(The restriction can be incorporated by 
a constraint in the action.)
We term this restricted theory
``minimal minimal Ho\v{r}ava gravity'', 
and denote it by the shorthand \mmHg.
It is equivalent to the Cuscuton model with quadratic potential for the scalar
Cuscuton field \cite{Afshordi:2006ad, Afshordi:2009tt,
Bhattacharyya:2016mah}, and to 
Einstein-aether theory with all aether couplings except $c_\theta$ set to zero,
restricted to the sector in which the aether expansion is nonzero
and the surfaces of constant expansion are spacelike \cite{Bhattacharyya:2016mah}.

The contribution of the
aether to the
metric field equation in {\mmHg} 
has the form of a perfect
fluid stress tensor with 
energy density
$\rho=-\frac{c_{\theta}}{6}K^2$ and isotropic pressure 
$p=\frac{c_{\theta}}{6}(K^2  -2\dot K)$,
where $\dot K$ denotes the derivative with respect to proper time along the unit
normal to the constant mean curvature (CMC) surfaces. The homogeneous isotropic case is simple to analyze, and allows for solutions that 
differ from those of general relativity. The effect of the aether stress tensor 
in the first Friedman equation
is to rescale the coefficient of the squared Hubble parameter (which is equivalent to rescaling the gravitational constant if the spatial curvature vanishes). This leads to the Big Bang Nucleosynthesis constraint,
$|c_\theta| \lesssim 0.3$ (assuming standard matter content) \cite{Carroll:2004ai}. It was argued in Ref.~\cite{Frusciante:2020gkx} that in Ho\v{r}ava gravity theory the CMB, combined with other cosmological observations, provides a
constraint an order of 
magnitude stronger than that,\footnote{It would be interesting to see if such detailed cosmology analysis might lead to improved bounds in the Einstein-aether theory.}
but is not clear that this analysis 
applies in the \mmHg\;limit 
of Ho\v{r}ava gravity. 

Another study 
with some observational bearing on 
\mmHg\;is that of Ref.~\cite{Wen:2023wes,Wen:2024orc},
which considers an effective dark energy 
fluid that becomes equivalent to \mmHg\;in the limit that the fluid sound speed 
goes to infinity~\cite{Afshordi:2006ad,Afshordi:2009tt}.
In that limit, their cosmic glitch parameter 
$\Omega_{\rm g}$ characterizing the deviation
from GR is equal to $-c_\theta/2$.
For practical reasons of numerical computation, however, this study analyzed the case with sound speed equal to the speed of light,
which might  produce results close to \mmHg, since other physics involved in structure formation involves much lower sound speeds. 
Fitting a number of 
cosmological observables 
(and assuming spatial flatness at large scales), 
they could impose
an upper limit $|\Omega_{\rm g}|<0.01$
at the 95\% confidence level,
but found evidence for a nonzero negative 
value of $\Omega_{\rm g}$.
Zero $\Omega_{\rm g}$ is disfavored 
at the $1.3\sigma$ -- $2.8\sigma$ level,  
depending on which large scale structure data are included.
It was found further that this 
model with nonzero negative
$\Omega_{\rm g}$
``alleviates both the Hubble and the clustering tensions
when considering the Planck 2018 data, 
while the $H_0$ measurement obtained under
Planck18 combined with DES Y1 data becomes more consistent with the SH0ES measurement.''
Although not
reported in ~\cite{Afshordi:2006ad,Afshordi:2009tt}, the authors found that taking the sound speed to be three times the speed of light decreased the best fit value for $|\Omega_{\rm g}|$ by about 27\%~\cite{Niayesh-personal}.
Whether a nonzero value would continue to be 
preferred when the sound speed is taken to infinity
remains unclear at present.
In any case, these results encourage a closer look at the foundations of the \mmHg\;theory.

To further diagnose the viability of the theory, it is important to 
see how it behaves when localized inhomogeneities develop. To gain some insight on this problem we analyze 
the case of collapse of a thin spherical shell. We find that the time evolution is unique and analytically computable up to the preferred hypersurface on which the shell contracts to a point\footnote{Note the distinction from the previously considered 
spherical collapse solutions \cite{Saravani:2013kva, relation}, 
in which the preferred
foliation was taken to 
consist of maximal slices. In that case the aether stress tensor vanishes, so the solution
is just that of general relativity with a `spectator'' aether. 
The preferred foliation limits to a universal horizon, stopping short of the singularity.}. Beyond this slice the solution is determined by a parameter with arbitrary time dependence, reflecting an indeterminacy of the theory in the presence of a singularity. This is a consequence of the infinite propagation speed inherent to theory, and is distinct from the familiar case of Einstein-Maxwell theory where the constraint of spherical symmetry is sufficient to guarantee a unique time-evolution beyond the Cauchy horizon of the Reissner-Nordstrom black hole. This 
non-uniqueness of evolution
once a singularity forms
seems to present a
serious challenge to the potential viability of the theory. 

It is an honor to dedicate this paper to the memory of Stanley Deser and 
his many contributions to the understanding of general relativity and its cousins.
Although Stanley would perhaps have had little patience for entertaining a Lorentz violating
cousin, particularly one that looks problematic from the get go, he did have an adventurous side of his own. Besides his work on foundational aspects of general relativity and supergravity,
he ventured, for example, into the worlds of 2+1 dimensions \cite{Deser:2019izm}, massive gravity \cite{Deser:2014fta}, and nonlocal cosmology \cite{Deser:2007jk,Deser:2013uya,Deser:2019lmm} (and references in all of these).
We imagine he would have both enjoyed our contribution, 
and put it in its place with a clever quip, as was his wont.

This paper is organized as follows.
Section 2 provides a definition of 
the IR limit of 
Ho\v{r}ava theory as a restricted 
version of Einstein-aether theory, 
and a broad orientation for the paper. 
Section 3 derives the equations of motion and conservation identities
for these theories. Section 4
examines the canonical structure 
of \mHg, and concludes 
that in order to have a chance to be viable
the theory must
be further restricted such that the
preferred foliation is a CMC foliation.
Section 5 first considers the homogeneous isotropic cosmological solutions, and then analyzes in detail 
the solution for a collapsing thin shell in such a cosmology. The key result is that, after the shell collapses to a singularity, the evolution becomes indeterminate. 
The conclusion summarizes the results
and briefly considers open questions.

\section{Ho\v{r}ava and Einstein-aether theories}

Ho\v{r}ava theory \cite{Horava:2009uw,Blas:2009qj}
is a deformation of general relativity 
with a preferred spacelike foliation of spacetime, and
is closely related to Einstein-aether theory \cite{Jacobson:2000xp} which  
has instead a preferred timelike threading of spacetime. It is
convenient for our purposes to view the IR limit of 
the former as a limiting case of the 
latter, so we will make use in this paper of both frameworks. 

Einstein-aether theory is a diffeomorphism invariant 
theory in which the metric of general relativity 
is coupled quadratically in (covariant) derivatives to a dynamical timelike 1-form  field of unit norm called the aether.\footnote{The aether is usually taken to be a vector; however, since the metric is assumed to be nondegenerate, the theory can equally well be formulated with a 1-form aether field. Use of the 1-form aether renders the relation 
to Ho\v{r}ava gravity more transparent.} 
The aether singles out a preferred timelike direction at any point in spacetime, breaking local Lorentz invariance in each configuration. 
The action for the IR limit of
Ho\v{r}ava gravity is identical \cite{Blas:2010hb}
to that for Einstein-aether theory with the
the aether 1-form restricted to be integrable --- 
equivalently, with the
aether vector  restricted to be orthogonal to the leaves of a foliation of spacetime, i.e., to be \textit{hypersurface orthogonal} (HO). This goes as follows:
\par
The general two-derivative action for Einstein-aether theory coupled to matter in $D+1$ dimensions can be written as
\begin{equation} \label{action}
    S=-\frac{1}{16\pi G_{\ae}}\int [R +\mathcal{L}_u]\sqrt{|g|}\,d^{D+1}x +S_{\rm m}
\end{equation}
where $\mathcal{L}_u$ is a sum of quadratic contractions of the aether $u_a$ and its covariant derivative, and the matter action is assumed to be 
independent of $u_a$.\footnote{The spacetime signature is timelike (mostly minus, 
$({+}{-}\cdots{-})$). The constant $G_{\ae}$ is neither 
the gravitational constant that appears in Newton's law nor the one in the cosmological Friedman equation---those receive (different) contributions from the aether couplings.
Note that $\mathcal{L}_u$ differs from the standard definition of Lagrangian density, because of the $-1/{16\pi G_{\ae}}$ prefactor.} 
It is natural \cite{Armendariz-Picon:2010nss, undoing} to decompose the derivative of the aether into tensors
that transform irreducibly 
with respect to the rotation group that preserves the aether,
\begin{equation}\label{irred}
    u_{a;b}=a_a u_b+\omega_{ab}+\sigma_{ab}-\tfrac{1}{D}\theta  h_{ab}\,.
\end{equation} 
Here $h_{ab}\equiv -g_{ab}+u_a u_b$ is the positive signature spatial metric orthogonal to 
the aether vector $u^a$. For future use we also define $h_a^b=\delta_{ab}-u_au^b$, the projector to the orthogonal subspace, and $h^{ab}=-g^{ab}+u^a u^b$ which inverts $h_{ab}$ on that subspace.

The four terms in \eqref{irred} can be identified with geometric properties of the
aether worldline congruence:
\begin{itemize}
    \item $a^a$ is the "acceleration", $u^a{}_{;b}u^b$\,,
    \item $\omega_{ab}$ is the "twist", the spatial projection of the antisymmetric part $h_a{}^m h_b{}^nu_{[n;m]}$\,,
    \item $\sigma_{ab}$ is the "shear", the traceless part of 
    the symmetric part $h_a{}^m h_b{}^nu_{(n;m)}$\,,
    \item $\theta$ is the "expansion", $u^a{}_{;a}$\,.
\end{itemize}
In terms of these quantities the aether 
``Lagrangian''
is 
\begin{equation}\label{Lu}
    \mathcal{L}_u = \tfrac{1}{D}c_\theta \theta^2+c_\sigma \sigma^2+c_a a^2+c_\omega \omega^2\,,
\end{equation} 
where the coefficients $c_{\theta,\sigma,a,\omega}$ are dimensionless
coupling constants.

In Ho\v{r}ava gravity the aether is restricted, at the level of the 
action, to be hypersurface orthogonal. By the Frobenius theorem this is equivalent to the vanishing of the twist three-form  $u\wedge du$, which
using the decomposition above is proportional to $\omega_{[ab}u_{c]}$. 
Hence the aether is HO if and only if $\omega_{ab}$ vanishes. In this case the other components also have natural interpretations on the leaves of the foliation. Since the aether covector is the unit normal, the projection of its covariant derivative onto a leaf is the extrinsic curvature $K_{ab}$ of that leaf, so $-\theta$ is the trace $K$ of the extrinsic curvature tensor and $\sigma_{ab}$ is the traceless part. Introducing a time function $T$ that labels the leaves of the foliation and writing $u=NdT$, where $N=1/|dT|$ is the ``lapse'', 
the acceleration is given by $a_a=-(\ln N)_{,b}h^b{}_a$. (Note that 
if $T$ is reparametrized, $N$ changes in such a way that $u_a$ is unaffected, so the action is invariant under $T$ reparametrizations.)
This enables us to rewrite the action in terms of $N$ and tensors on leaves
of the foliation as
\begin{equation}\label{action2}
 S =\frac{1}{16\pi G_{\ae}}\int dT d^Dx \,N\sqrt{h}\bigl[(1-c_\sigma)K_{ij}K^{ij}-\lambda K^2 + {}^{(D)}\!R +c_a a_ia^i\bigr] 
 \end{equation}
where $\lambda = 1+\frac{1}{D}(c_\theta-c_\sigma)$, and the letters $i,j$ from the middle of the latin alphabet are spatial indices, i.e., indices for tensors on the leaves of the $T$-foliation. 
This is precisely the form of the IR limit of the Ho\v{r}ava-Lifshitz action, in $D$ spatial dimensions~\cite{Horava:2009uw}. 
Notably, it is invariant under \textit{foliation-preserving} diffeomorphisms, i.e. time-dependent spatial diffeomorphisms and reparametrizations of the time coordinate $T$. 

  The parameter space for the coupling constants is heavily restricted by 
  stability of the theory (real wave frequencies and positive energy), 
  and by observations.  For  recent compilations of constraints see~\cite{EmirGumrukcuoglu:2017cfa, Frusciante:2020gkx}.
For comparison, we note that the parameters used in \cite{relation} match ours by 
 \begin{equation}
     \alpha=c_a,\;\beta=c_\sigma,\; \lambda'=\frac{1}{D}(c_\theta-c_\sigma)
 \end{equation}
 where $\lambda'$ is what those authors denote as $\lambda$.
  The strongest observational constraints  
  come from the absence of dissipation of cosmic ray energy
  to \v{C}erenkov radiation into the spin-2 and spin-0 modes (which implies that those
  wave speeds are greater than or equal to the speed of light for matter) and from
  comparison of the speeds of gravitational and electromagnetic 
  waves from the neutron star collision GW170817, which sets a bound  
  $|\beta|\lesssim 10^{-15}$. 
  The next strongest constraints arise from 
  solar system observations, and from orbital dynamics of binary and ternary pulsars.
  The resulting allowed parameter regions are
 either $|\alpha|\leq10^{-7}$ or $|\alpha|\leq10^{-4},\lambda'\leq\frac{\alpha}{1-2\alpha}$.
There are also the BBN and cosmological constraints on $c_\theta$ mentioned in the Introduction. 
 
The focus of this paper is on the case with only $c_\theta$ being nonzero, called \textit{minimal} Ho\v{r}ava gravity (\mHg)~\cite{minimal}.
 Existing work has shown that, for asymptotically flat data, 
this special case is equivalent to vacuum GR in maximal gauge, i.e., with
$K=0$ on all slices~\cite{consistency}. 
 In this study we establish that allowing for general boundary conditions, 
 and discarding some pathological solutions, the $K=0$ constraint is relaxed 
to permit time-varying, spatially constant $K$, where the value of $K$ on each 
time slice is determined by an elliptic equation.
The aether stress tensor in this case is nonzero, and takes the form of a perfect fluid
with a time-dependent equation of state. 
The problem of spherical collapse of a thin shell with asymptotic cosmological boundary 
conditions is analyzed, and the time evolution is found to become ambiguous past the preferred slice on which the shell contracts to a point.
This exhibits a markedly different qualitative behaviour from the previously considered maximally sliced solutions~\cite{relation}, 
where the maximal foliation on which the \mHg\;solution is defined stops before reaching 
the singularity that would appear in GR.

 \section{Field equations, constraints, and conservation identities}
 In this section we spell out the field equations for both Einstein-aether theory 
 and for the IR-limit of Ho\v{r}ava theory, exhibit their relation, and identify the 
 constraints and conservation identities that follow from diffeomorphism invariance. 
 
 \subsection{Field equations}
Restricting the action \ref{action} to the space of hypersurface orthogonal configurations and imposing stationarity produces the equations of motion of the IR-limit of Ho\v{r}ava theory in diffeomorphism invariant form, a description that has been referred to as  $T$-theory~\cite{extended}, or more commonly khronometric theory~\cite{relation}. 
As we shall see, the field equation resulting from the variation of $T$ is automatically satisfied if the other field equations are~\cite{extended}.
However, this implication does not hold separately at each time, since
both the metric field equation and its time derivative at one time are required 
to infer the $T$ field equation at that time. 
Note that these equations are \textit{not} the same as those of Einstein-aether theory; the action in khronometric theory is not required to be stationary under variations that break hypersurface orthogonality. In particular, Einstein-aether theory includes an additional $D$ equations per spacetime point, arising from variations of the aether satisfying the unit norm constraint~\cite{two_versions}. 
All hypersurface orthogonal solutions of Einstein-aether theory are solutions of khronometric theory, but solutions of khronometric theory are in general not solutions 
of Einstein-aether theory.
\par

Let us begin with the case of Einstein-aether theory.
The constraint that the aether is a unit vector can be imposed with a Lagrange multiplier term in the action, or just by restricting the variations to those that preserve the unit norm condition $u_a u^a = g^{ab}u_au_b=1$. Adopting the latter 
method, it is useful to split the variations into two classes $\delta_{\parallel}$ and $\delta_{\perp}$ as follows:
 \begin{align}
\delta_{\parallel}g_{ab}&=\delta g_{ab},\quad\delta_{\parallel} u_{c}=\frac{1}{2}(\delta g_{ab}\,u^a u^b)u_c\\
\delta_{\perp}g_{ab}&=0,\quad u^c\,\delta_{\perp}u_c=0
 \end{align}
 The parallel variation $\delta_{\parallel}$ varies the metric, and performs the required pure scaling of the \textit{covector} aether $u_a$ to maintain the unit norm constraint, while the perpendicular variation $\delta_{\perp}$ varies only $u_a$ in a direction orthogonal to itself, and thus automatically maintains the unit norm constraint to first order. We can define an aether ``stress tensor'' $T^u_{ab}$, capturing the effect of
 the full parallel variation, via 
\begin{align}
    \dpar S_u &=:\int  \frac12 T^u_{ab}\,\delta g^{ab} \sqrt{|g|}\,d^{D+1}x\,,\label{Tu}\\
    &=
    \int \Bigl(\frac12 {}^{(g)}T^u_{ab} + \frac12\AE^{c}u_c u_a u_b\Bigr)\delta g^{ab} \sqrt{|g|}\,d^{D+1}x
\end{align}
where $S_u$ is the aether contribution to the action \eqref{action}, and 
\begin{equation}\label{TAE}
{}^{(g)}T^u_{ab}:= \frac{2}{\sqrt{|g|}}\frac{\delta S_u}{\delta g^{ab}}
\qquad\mbox{and}\qquad
\AE^{c}:= \frac{\delta S_u}{\delta u_{c}}\,.
\end{equation}
Note that 
$\AE^b$ is a vector density of weight one, as 
it arises from variational derivative of the action with respect to a covector.
The full set of 
equations of motion arises from setting all variations to zero, 
which yields an Einstein equation from the parallel variation, together 
with the aether equation from the perpendicular variation: 
\begin{align}\label{einstein}
     G_{ab}&=T^u_{ab}+ T^{\rm m}_{ab}\qquad\mbox{and}\qquad
     (\AE^b)^\perp=\AE^a h_a{}^b =0\,. 
 \end{align}
Here and in the following we have adopted units with $8\pi G_{\ae}=1$.

In khronometric theory the aether is constructed as a unit vector from $T$ and $g_{ab}$, 
\begin{equation}\label{u=dT/|T|}
 u_a:=T_{,a}/\sqrt{g^{mn} T_{,m}T_{,n}}\,,   
\end{equation}
and the equations of motion arise from independent variations of $g_{ab}$ and $T$.  
The variation of $g_{ab}$ alone induces a parallel variation of the aether 
\eqref{u=dT/|T|}
that preserves its unit norm, so corresponds to the previously defined parallel variation and thus leads to an Einstein equation involving the effective stress tensor $T_u^{ab}$ defined 
above in \eqref{Tu}. 
The variation of $T$ alone induces a perpendicular variation of the aether,
\begin{equation}
    \delta_T u_a = N h_a{}^b (\delta T)_{,b}\,,
\end{equation}
and the corresponding variation of the action is
\begin{equation}
    \delta_T S = \int \AE^a  N h_a{}^b (\delta T)_{,b} 
    = -\int  (N \AE^a h_a{}^b)_{,b} \delta T\,.  
\end{equation}
The $T$ field equation is thus given by\footnote{The coordinate divergence of a vector density of weight one is a scalar density of weight one, and is equal to the covariant divergence in 
local inertial coordiantes at each point, hence these two divergences are different ways of writing the same scalar density.}  
\begin{equation}\label{Teqn}
 \mathcal{T} := (N\AE^a h_a{}^b)_{,b}=0\,.
\end{equation}
This exhibits explicitly how the perpendicular aether field equation implies
the $T$ field equation, but not vice versa.\footnote{In the special case of 
spherical symmetry with a regular center, 
the radial component $\AE^r$
must vanish at $r=0$, in which case 
\eqref{Teqn} implies the radial component---and thus the full---perpendicular aether field equation, since we assume that $N\ne0$.
(A different argument for this was given in Appendix D of \cite{Blas:2010hb}.)}

\subsection{Constraints and conservation identities in khronometric theory}\label{sec:conservation}
\label{ConsId}

Diffeomorphism symmetry is responsible for the presence of initial value constraints among
the field equations, as well as conservation laws relating to the preservation of the
constraints. In this section we spell out these features of the theory, as they 
are important for understanding the structure of \mHg\;as well as for our construction
of a cosmological solution with a collapsing shell of matter.

 The variation of the khronometric theory action 
 under a diffeomorphism induced by a vector field $\xi^a$ is 
 \begin{align}\label{T-diffeo2}
    \delta_\xi S[g_{ab}, T] &= \int E^{ab}\,\delta_\xi g_{ab} +\mathcal{T}\, \delta_\xi T\\
    &=\int 2E^{ab}\,\xi_{a;b} + \mathcal{T} T_{,a}\, \xi^a\\
    &=\int (-2E_a{}^b{}_{;b}+ \mathcal{T} T_{,a})\, \xi^a
\end{align}
Here $E^{ab}$ and $\mathcal{T}$  are the variational derivatives 
with respect to $g_{ab}$ and $T$, respectively, and 
it is assumed that $\xi^a$ vanishes at the boundary so no boundary term appears. 
Diffeomorphism invariance of the action thus implies the {\it identity}  
\begin{equation}\label{diffid}
    2E_a{}^b{}_{;b}=\mathcal{T} T_{,a}
\end{equation}
From this 
equation it follows that if the metric field equation ($E^{ab}=0$) holds everywhere, and if $T_{,a}\ne0$ (which is required in order for the level sets of $T$ to define a nondegenerate foliation of the spacetime), then the $T$ field equation ($\mathcal{T}=0$) holds everywhere.
In that sense, the $T$ field equation need not be imposed separately. 
(Note, however, that the metric field equation on a single time slice does {\it not}
imply the $T$ field equation on that time slice, since \eqref{diffid} involves the time derivative of the metric field equation.)
Conversely, if $\mathcal{T}=0$ holds then $(E_a{}^b{}_{;b})_{\parallel}=0$, i.e., $u^a E_a{}^b{}_{;b}=0$.
Moreover, since $T_{,a}$ has no perpendicular component, it follows independently of any field equations that $(E_a{}^b{}_{;b})_{\perp}=0$. Since the Einstein tensor term in $E_a{}{}^b$ has identically vanishing divergence, these relations apply also to the aether stress tensor alone.

The divergence equation $E_a{}^b{}_{;b}=0$ implies that, for any coordinate $x^0$, the
derivative $\partial_0 E_a{}^0$ vanishes at a value of $x^0$ if the field equations
$E_a{}^b=0$ hold at that value of $x^0$. 
In khronometric theory, 
when $x^0 = T$, the $T$ field equation thus implies that all four of the 
$T$-component metric equations, $E_a{}^T=0$, are preserved in time (but note that 
the three perpendicular 
equations are preserved independently of the $T$ equation).
The significance of this observation is that these equations 
are initial value constraints with respect to $T$ evolution, 
i.e., they involve no second or higher order derivatives with respect to $T$.
(See \cite{initial} for a detailed exposition of the constraint analysis.)
The $T$ equation thus ensures preservation of the $u$-component of the constraints.

\subsection{Minimal Ho\v{r}ava Gravity}
\label{mHg}
 From here on out we restrict to the minimal Ho\v{r}ava case, where the only
 nonzero coupling in the aether Lagrangian \eqref{Lu} is $c_\theta$, i.e., only the
 expansion term is present. In this case, $\AE^a\equiv \delta S/\delta u_a$ is given by 
\begin{equation}\label{ae_eqn}
    \AE^a = -\frac{c_\theta}{D} \sqrt{|g|} g^{ab}\theta_{,b}\,.
\end{equation}
In Einstein-aether theory, the aether field equation \eqref{einstein} 
imposes that
this has no component perpendicular to the aether, i.e., 
$\AE^a = \AE^b u_b\, u^a$, hence $\theta_{,a} =\dot\theta u_a$,
where $\dot \theta \equiv u^b \theta_{,b}$. If $\dot\theta\ne0$ this 
implies that $u^a$ is orthogonal to surfaces of constant $\theta$, 
while if $\dot\theta=0$ it implies that $\theta$ is constant in spacetime.\footnote{In the latter case, the aether stress tensor is just that of a cosmological constant, but as noted in \cite{Bhattacharyya:2016mah} the aether vector evolution is underdetermined.}
In Khronometric theory,the $T$ field equation \eqref{Teqn} takes the form  
\begin{equation}\label{mTeqn}
    \mathcal{T} = -\frac{c_\theta}{D} (N^2 \sqrt{h} h^{ab}\theta_{,a})_{,b}=0\,.
\end{equation}
This equation is satisfied if $\theta$ 
is constant on the leaves of the $T$ foliation, i.e., if it is a constant mean curvature
(CMC) foliation. Otherwise, the equation imposes 
a relation between the lapse $N$ and $\theta$. 

Locally it appears that either of these ways of satisfying \eqref{mTeqn}
are possible; however, under some boundary conditions \eqref{mTeqn}
actually implies that $h^{ab}\theta_{,a}=0$. For example, if the spatial
slices are compact without boundary, one can 
multiply \eqref{mTeqn} by $\theta$, integrate over the spatial slice, 
and integrate by parts, yielding
the integral of the positive definite 
quantity $N^2 \sqrt{h} h^{ab}\theta_{,a}\theta_{,b}$, from which it
follows that $h^{ab}\theta_{,a}=0$, assuming that $N\ne0$ (see \cite{consistency}
for a Hamiltonian version of this argument). 
Alternatively,
if one restricts to asymptotically flat spatial slices in asymptotically
flat spacetime, a similar argument can be made, and in that case one
arrives at the stronger restriction that the foliation is maximal (i.e., of zero mean curvature). 
In that case the aether stress tensor \eqref{TTT} vanishes, 
hence the theory is equivalent to general relativity 
in a maximal slicing gauge, if such exists \cite{consistency}.
In the present paper we shall consider noncompact spatial slices
in a spacetime that is not asymptotically flat, so will need to contend with the 
fact that there are two distinct ways to satisfy \eqref{mTeqn}.
As we'll see, to avoid dynamical pathology in that case it is necessary
to impose the extra condition that \eqref{mTeqn} be satisfied by virtue of
$h^{ab}\theta_{,a}=0$.

The stress tensor  \eqref{Tu} for the aether is given by
\begin{equation}\label{TTT}
    T^u_{ab} = \frac{c_\theta}{D}\left(
    -\tfrac12\theta^2 g_{ab} + \dot\theta h_{ab} + 2 \theta_{,m} h^m{}_{(a} u_{b)}\right)\,.
\end{equation}
The first term has the form of a time-dependent cosmological constant, the second term is an isotropic pressure term, and the third term is an energy flux/momentum density in the direction of the gradient of $\theta$ along the leaves of the $T$ foliation. Note that if $\theta$ is constant everywhere in spacetime, then the stress tensor reduces to that of a cosmological constant, so the equations of motion reduce to the Einstein equation with a cosmological constant. If $\theta$ is constant on the leaves of the $T$ foliation, then the stress tensor takes the form of a perfect fluid with an evolving equation of state.

\section{Initial Value Formalism}
\subsection{Canonical Structure}
Our next task is to understand the nature of the propagating degrees of freedom, which is best done in the canonical formalism. This looks somewhat different from that developed in \cite{hamiltonian_structure} because of the vanishing of $c_a$,
which means that the lapse contributes to the action only as a Lagrange multiplier. 
Our implementation will instead follow more closely \cite{hamiltonian_dynamics} after dropping higher derivative terms (see also \cite{consistency,minimal}). We will formulate the Hamiltonian structure in a gauge adapted to the preferred slices, establish the Lagrange multiplier equations which supplement the Einstein equation (with the stress tensor calculated in the previous section) and the matter equations of motion, and count the degrees of freedom at the linearized level. This will reveal exotic features of \mHg\;under arbitrary boundary conditions.

We begin with the 3+1 ADM decomposition \cite{Arnowitt:1962hi} of the action \eqref{action2}, setting
$c_\sigma=c_a=0$:
\begin{equation}\label{actionm}
 S_{\rm mHg} =\frac{1}{16\pi G_{\ae}}\int dT d^Dx \,N\sqrt{h}\bigl(K_{ij}K^{ij}-\lambda K^2 + {}^{(D)}\!R\bigr) 
 \end{equation}
The spacetime metric is decomposed as
\begin{equation}
    ds^2 = N^2 dT^2 - h_{ij}(dx^i + N^i dT)(dx^j + N^j dT)\,.
\end{equation}
The configuration coordinates in the canonical formalism are the 
spatial metric $h_{ij}$ on surfaces of constant $T$, while 
the lapse $N$ and shift $N^i$ are Lagrange multipliers. 
The conjugate momentum is 
\begin{equation}
      \pi^{ij}=\sqrt{h}(K_{ij}-\lambda K h_{ij})
\end{equation}
and the Hamiltonian is \cite{consistency,hamiltonian_dynamics}
\begin{equation}\label{Hamiltonian}
    H=\int(N\mathcal{H}+N^i\mathcal{H}_i) d^3x, 
    \end{equation}
 with
 \begin{align}
 \mathcal{H}&=\frac{1}{\sqrt{h}}\left(\pi^{ij}\pi_{ij}-\frac{\lambda}{D\lambda-1}\pi^2\right)+\sqrt{h}\mathcal{R}\\
 \mathcal{H}^i&=-2D_j\pi^{ij},
\end{align}
where indices are raised and lowered by the spatial metric, $D_i$ is the spatial 
covariant derivative, and $\mathcal{R}$ is the spatial Ricci scalar. 
The variational equations for the lapse and shift imply initial data constraints,
i.e., the vanishing of 
$\h$ and $\h_i$, which are the generators of perpendicular evolution (Hamiltonian density) and spatial diffeomorphisms (momentum density), respectively, provided that $\lambda$ is not at the singular point $\frac{1}{D}$ (which we assume from here on). 
If matter is coupled minimally to the metric and not coupled to the $T$ field 
the structure of the Hamiltonian remains the same as \eqref{Hamiltonian}, 
with matter terms $\h_{\rm m}$ and $\h^i_{\rm m}$ added to the constraints. 
The Poisson brackets of the constraints are given by 
\begin{align}\label{algebra}
\{\h_i(x),\h_j(y)\}&=\h_i(y)\partial_j^x\delta(x,y)+\h_j(x)\partial_i^x\delta(x,y)\nonumber
\\
\{\h(x),\h_i(y)\}&=-\h(y)\partial_i^y\delta(x,y)\nonumber
\\
    \{\mathcal{H}(x),\mathcal{H}(y)\}&=(\h^i(x)+C^i(x))\partial^x_i\delta(x,y)-(\mathcal{H}^j(y)+C^j(y))\partial^y_j\delta(x,y),
    \end{align}
 where 
 \begin{equation}\label{ci}
     C^i(x):=2\frac{\lambda-1}{D\lambda-1}D^i\pi\,.
 \end{equation}
The first two brackets are identical to those of general relativity;  they just express the effect of an infinitesimal spatial diffeomorphism 
on the vector density and scalar density constraints. 
For the same reason, they also hold if matter contributions are included in the constraints. Under the minimal coupling assumption about the matter, the third bracket 
can be split into the sum of the brackets among the matter and gravity parts separately,
the matter part taking the standard form since the only nonzero brackets that arise are then from the matter fields. The gravity part 
does \textit{not} take the same form as in GR --- there is an additional contribution involving $C^i$. 
As shown by the classic work of Hojman, Kuchar, and Teitelboim \cite{geometrodynamics}, 
the presence of the $C^i$ term means that the hypersurface deformation algebra does not close, which implies that one cannot consistently evolve the data on the initial slice according to an arbitrary lapse, as evolving along different lapses to the same spatial slice would produce inequivalent results. This is to be expected, since the theory has a preferred foliation, that of the constant $T$ surfaces, and consistent Hamiltonian evolution must involve a lapse that preserves this foliation.
Therefore, we expect that the constraints will generate restrictions on the lapse.

The $T$ derivative of any observable ${\cal O}$
is given by $\dot {\cal O}\equiv \partial_T {\cal O} = \{{\cal O},H\}$, 
where $H$ is the Hamiltonian \eqref{Hamiltonian}. 
Using the bracket algebra \eqref{algebra} one obtains the time derivatives of the constraints,
\begin{align}
    \dot{\h_i}&=\h N_{,i}+ D_j(N^j \h_i)+\h_jD_iN^j\\ 
    \dot{\h}&=(N^i\h)_{,i}+2N_{,i}(C^i+\h^i)+N(\h^i+C^i)_{,i}\,.
\end{align}
The preservation of the momentum constraint for arbitrary lapse is automatic as long as the constraints hold initially.\footnote{If only the momentum constraint is imposed, its preservation is guaranteed if the lapse is constant. The resulting theory is referred to as \textit{projectable} Ho\v{r}ava gravity. 
We are interested here in the non-projectable case.} The preservation of the Hamiltonian constraint requires in addition that
\begin{equation}\label{T-ham}
    (N^2C^i)_{,i}=0\,.
\end{equation} 
It is no accident that this is equivalent to the $T$ field equation \eqref{mTeqn}; as explained in section \ref{ConsId}, the $T$ field equation 
is equivalent to the preservation of the $T$-component of the 
constraints. 
Equation \eqref{T-ham} must be imposed at each time.
As already discussed in the paragraph under \eqref{mTeqn} 
in the context of the covariant formulation, the local character of the solutions of this system of equations
has a strange dichotomy. Equation \eqref{T-ham} 
can be satisfied in two quite different ways. In case (A), the vanishing of 
$C^i$---i.e., the vanishing of $K_{,i}$---implies that the Hamiltonian constraint $\mathcal{H}$ is first class. Therefore
\eqref{T-ham} is satisfied without regard to the lapse function $N$.
Note that $C^i=0$ is equivalent to the aether equations \eqref{einstein} 
for the special case of \mHg.
In case (B), \eqref{T-ham} imposes a nontrivial restriction on $N$, which determines $N$ from 
data on a codimension-1 submanifold of the spatial slice. It is even possible for both situations to occur in different regions of spacetime, or on the same spatial slice. 

Case (B) 
exhibits a pathological "half-mode"~\cite{Li:2009bg,Blas:2009yd}.
This arises because the momentum constraints are all first class, but the Hamiltonian constraint is second class without a "partner" constraint; it fails to Poisson commute with itself at different points on the slice. Dirac's heuristic prescription for counting degrees of freedom (per space point) 
tells us that each second class constraint eliminates one free canonical variable, but a first class constraint eliminates two, because in addition to constraining initial data it generates a gauge transformation which parametrizes redundancy of the description. The number of degrees of freedom is half the number of remaining canonical variables. Applying this formula to case (B), we find $\frac{1}{2}[D(D+1)-2D-1]$ which is a half integer. For comparison, in GR the Hamiltonian constraint is also first class, so the number of degrees of freedom is $\frac{1}{2}[D(D+1)-2D-2]$, which is $\frac{1}{2}$ less, and thus a whole integer.

The nature of the half-mode was studied in \cite{Blas:2009yd},
where it was shown that, on a generic background, it satisfies 
an equation first order in time, and is therefore specified by only one real function (as opposed to ordinary modes, which are specified by
two functions of initial data). Moreover, 
on a patch of spacetime on which the background field coefficients 
that enter the linearized equation (for example $K_{,i}$) 
are translation invariant, 
the half-mode (for wavelengths much smaller than the patch size) 
has a peculiar, background-dependent  dispersion relation,
and is unstable at high frequencies and strongly 
coupled at low frequencies. No such mode has been observed in Nature, 
and Case (B) must therefore be excluded in a viable physical theory.

To rescue the theory from the oblivion of case (B), we shall restrict the theory to case (A), i.e., we require that the mean curvature $K$ of each spatial surface of constant $T$ be constant on that surface. The constant $T$ surfaces thus provide a CMC foliation for the spacetime. For general boundary conditions,
This restriction amounts to a change of the definition of the theory under consideration. However, as discussed in the covariant setting above,
under some boundary conditions it is anyway implied by \eqref{T-ham}.
If the spatial slices are compact without boundary, then 
\eqref{T-ham} actually implies that $K_{,i}=0$~\cite{consistency}. 
The same is true if
the spacetime and foliation are asymptotically flat, in which case 
the stronger condition $K=0$ follows and resulting theory is then nothing
but general relativity expressed in maximal slicing gauge \cite{consistency}. 

In a general setting, however, $K$ evolves on
the preferred foliation, 
so even if $K_{,i}=0$ 
the theory differs in some peculiar nonlocal  
fashion from general relativity due to the aether contribution to the stress tensor. 
As mentioned in the introduction,
the motivation for the present paper is to explore the nature 
and viability of this peculiar deformation of GR.

A convenient way to impose the CMC condition on the theory is to specify $K = K(T)$ 
as a function of $T$. This eliminates the half-mode and, provided the function is invertible, also gauge fixes the time reparametrization symmetry.\footnote{The 
theory with this constraint 
can be characterized in the covariant formulation (a.k.a.\ Stueckelberg formulation \cite{Blas:2009yd}) by adding to the Lagrangian a term 
$\alpha(\theta + T)$, where $\alpha$ is a Lagrange multiplier.} 
The number of degrees of freedom per point in this restricted theory are
the same as in GR, so it is what is known as a minimally modified model of gravity~\cite{relation}.
This is why we refer to it as \textit{minimal} 
minimal Ho\v{r}ava gravity \mmHg\;for short).\footnote{As discussed in section \ref{mHg}, for 
spacetimes with 
closed or asymptotically flat spatial slices, \mmHg ${}={}$\mHg.}
For the remainder of this paper we focus solely on this case.

\subsection{Cauchy Problem}\label{sec:cauchy}
The equations of motion are precisely the corresponding equations for GR, coupled to the aether stress tensor \eqref{TTT} as well as any matter terms.
There is also a condition arising from requiring the mean curvature $K$ to be a prescribed function of time, which eliminates the spatial gradient term in the aether stress
tensor \eqref{TTT}.
Under time reparametrizations $K$ transforms as a scalar, and fixing its value at each time provides a choice of gauge. If $K$ is chosen to be a constant on a given time interval, then the vanishing of the $\dot{K}$ term in \eqref{TTT} implies that the solution is also a solution of GR coupled to the same matter fields, but with an addition of  $\frac{c_\theta}{D}\frac{1}{2}K^2$ to the cosmological constant. The other prototypical possibility is that $K$ is a strictly monotonic function over a particular time interval, in which case it can be used as the time coordinate. Such regimes are joined to each other at points where $\dot{K}=0$.
\par
The equation governing the evolution of $K$ can be derived from the Einstein evolution equations. 
The resulting equation (under the assumption of shiftless gauge)
is 
\begin{align}
    \partial_t K &= \left[-\Delta+ K_{ij}K^{ij}+\frac{1}{2}T_{\rm tot}\right]N\\
    &=\left[-\Delta+ K_{ij}K^{ij}+\frac{c_\theta}{2D}\left(\frac{D\cdot\partial _t K}{N}-\frac{D+1}{2}K^2\right)+\frac{1}{2}T_{\rm m}\right]N
\end{align}
where $T_{\rm tot}$ is the spacetime trace of the total (aether plus matter) stress tensor.
(Note that we adopt from now on the letter $t$ to denote the
preferred time coordinate $T$ of the Ho\v{r}ava theory, to avoid confusion with the
stress tensor.)
This results in a linear elliptic PDE,
\begin{equation}\label{lapse}
    \partial_t K=L N := \frac{1}{1-{c_\theta}/{2}}\left[-\Delta + K_{ij}K^{ij}-\frac{c_\theta(D+1)}{4D}K^2+\frac{1}{2}T_{\rm m}\right]N\,.
\end{equation}
In the first case, where $K$ is a constant over a time interval, the LHS is set to 0 and the homogeneous boundary value problem 
for $L$ must be solved to find $N$ on each time slice. This solution must then be combined with the other evolution equations to evolve the metric and extrinsic curvature forwards in time. As discussed above, this essentially amounts to solving GR with a cosmological constant (arising from the $\theta^2$ term in \eqref{TTT} together with 
any contribution from $T_{\rm m}^{ab}$) in CMC gauge.
On the other hand, in the case where $K$ is a monotonic 
function that can act as a time coordinate, $\partial_t K$ is set to 1, giving a particular inhomogeneous problem on each time slice. The Fredholm alternative tells us that \textit{either} $L$ has a zero mode \textit{or} it is invertible, so if there is no solution to the homogeneous boundary value problem, there is a unique solution where $\partial_t K=1$. In fact, since the existence of a zero mode is a strong condition on the
spectrum of $L$ if the spatial slice is compact, the constant $K$ regime is not viable in this case. Although we are not restricting to compact spatial slices, we will still restrict to the monotonic case for the remainder of this paper to explore behaviour that is distinct from that of GR.
Note that we cannot guarantee that this Cauchy problem is well-posed for an arbitrarily long interval of CMC time, since the existence of a solution is essentially a driven version of the as-yet unresolved CMC conjecture in GR, which states that a CMC Cauchy slice in a vacuum spacetime is part of a foliation that extends for infinite CMC time in at least one direction  \cite{cmc_gauge}. If time evolution indeed fails to be solvable at some time, this represents a breakdown of the theory.

Furthermore, the ellipticity of the operator $L$ implies that even when the evolution does exist, it is not causal. While GR in CMC gauge obeys a similar equation, the difference is that there it is a gauge condition, which can be replaced by a manifestly causal wave gauge. This is not the case in \mHg, where the slicing actually influences the stress tensor. When $K$ is strictly monotonic, the stress tensor clearly depends explicitly on the preferred slicing, and will be affected by spacelike separated disturbances on an initial slice which distort the evolution of that slicing via the lapse equation. Therefore the theory has a physically meaningful infinite propagation speed.

\section{Spherically Symmetric Solutions}

The questions of whether the initial value problem is well-posed 
in the theory, and of whether dynamical solutions are 
stable despite the infinite propagation speed,
are clearly important ones that would have to be addressed and settled 
in the affirmative if the theory were to be a viable candidate for 
describing Nature. We shall not attempt to tackle these
questions here, but instead just explore some particular 
spherically symmetric solutions, 
in order to expose some features of the theory.
From here on we restrict to the case of $D=3$ spatial dimensions.

\subsection{Cosmological Spacetimes}

We first consider solutions to $\mathrm{m^2Hg}$ for which the spacetime and aether are homogeneous and isotropic, with no additional symmetry (note that this rules out exact de Sitter).
The aether is then orthogonal to the maximally symmetric cosmological hypersurfaces, and 
automatically satisfies its field equation, so only the 
metric field equation need be imposed. This
is the same in $\mathrm{m^2Hg}$ as in Einstein-aether theory.
With this symmetry, even if the shear, twist, and acceleration
terms are present in the Lagrangian, 
the aether stress tensor receives contributions only from the 
expansion term, and is given by \eqref{TTT} without the spatial
gradient,
\begin{equation}\label{TTT2}
    T^u_{ab} = -\frac{c_\theta}{3}\left(
    \frac12\theta^2 g_{ab} - \dot\theta h_{ab}\right)\,.
\end{equation}
This has the form of an isotropic perfect fluid.\footnote{In the special case 
when all leaves of the spatial foliation have the same
extrinsic curvature $K=-\theta$, 
the aether stress tensor \eqref{TTT}
takes the form of a cosmological constant term, 
hence such geometries can be solutions 
in both GR with a cosmological 
constant and in Ho\v{r}ava and Einstein-aether theory.} 
Since the aether automatically
satisfies its field equation, its stress tensor is identically conserved, 
and thus it must be a linear combination of the Einstein tensor $G_{ab}$ and
the spatial curvature fluid; in fact, it takes the form
\begin{equation}\label{Friedmann}
    T_{ab}^u=-\frac{c_\theta}{2}\left(G_{ab}-\frac{1}{6}\mathcal{R}(g_{ab}+2u_a u_b)\right)
\end{equation}
where ${\cal R}$ is the spatial curvature \cite{aether}.\footnote{With FLRW symmetry, the divergence of the ${\cal R}$ term in \eqref{Friedmann}
is proportional to $\dot{\cal R} + 2 {\cal R} \dot a/a$ (where $a$ is the scale factor), 
which vanishes since ${\cal R}\propto a^{-2}$.}
If ${\cal R}$ vanishes
on the preferred foliation,
as is seemingly the case 
on cosmological scales,\footnote{See, however, \cite{Yang:2022kho}.}
$T^u_{ab}$ is proportional to the Einstein tensor, 
\begin{equation}\label{TTT3}
    T^u_{ab} = -\frac{c_\theta}{2}G_{ab}\,,
\end{equation}
so it can be absorbed into a redefinition of the gravitational constant $G_{\rm cosmo}$
appearing
in the Friedman equations, $G_{\rm cosmo} = (1+{c_\theta}/{2})^{-1} G_{\ae}$.
As discussed in the
Introduction, the difference between
$G_{\rm cosmo}$ and the gravitational constant relevant for
local dynamics, $G_{\rm N}= (1-{c_a}/{2})^{-1}G_{\ae}$,
has observable effects on Big Bang Nucleosynthesis 
and large scale structure.

To probe the peculiar, signature features of the theory, we turn attention now to the example of 
spherically symmetric collapse of a shell of matter.
This serves as a primitive example of the sort of process that gives rise to nonlinear
structure in the universe when small deviations from maximal spatial symmetry
grow due to their self-gravity, leading eventually to the formation of stars, 
galaxies, and black holes.

\subsection{Spherical Collapse}
\begin{figure}
     \centering 
     \includegraphics[width=0.4\linewidth]{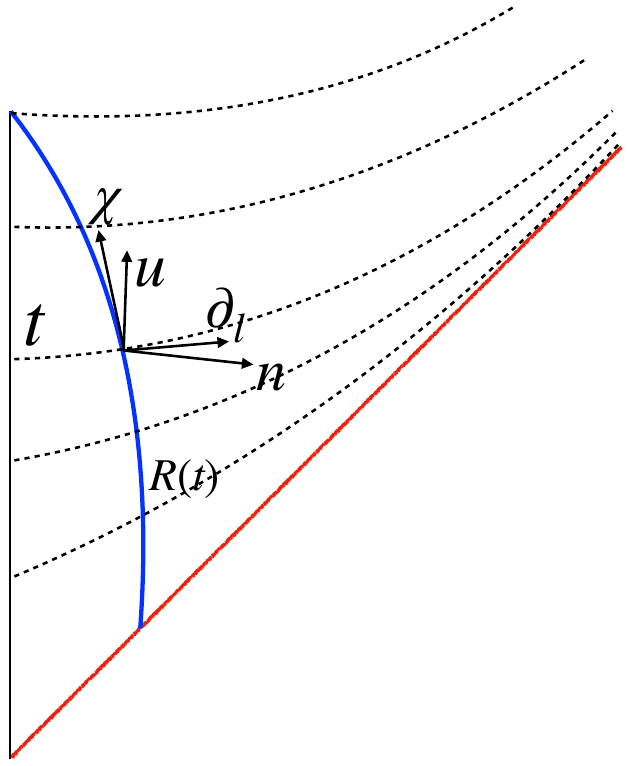}
     \caption{Spacetime diagram of collapsing spherical shell. The blue curve is the shell of spherical radius $R(t)$, where $t=-3/K$ labels the surface of constant mean curvature $K$ (dotted curves). Inside the shell is the Milne-like geometry \eqref{intds2}, which is singular on the red line. (The singularity is presumed to continue as null outside the shell.) 
     $\chi$ is the 4-velocity of the shell and $n$ is the outward unit normal; 
     $u$ is the aether vector and $\partial_l$ is the unit radial tangent 
     to the constant $t$ surface. 
     \label{fig:sketch}}
 \end{figure}

Consider the collapse of a freely falling thin spherical shell of pressureless dust.\footnote{We will take the shell to be of infinitesimal thickness, but we expect this to be a good approximation for a shell of finite but small thickness as far as the qualitative behavior of the solution is concerned.}
The inside is taken to be an empty Friedmann universe, and the exterior is a vacuum \mmHg\;geometry that we seek to solve for. The two are joined across a boundary with the appropriate surface stress-energy tensor, requiring the metric to be continuous across the boundary as a tensor field. A spacetime sketch of the sort of solution 
 we consider, along with some quantities used in the following analysis, is shown in Fig.~\ref{fig:sketch}.

 The areal radius of the shell is described by a function $R(t)$; from this, 
 together with the spacetime metric on either side, the 4-velocity of the shell $\chi$ can be computed (and the continuity of the metric implies that it does not depend on which side 
 is used).
 We will find that the constraint equations on the CMC slices fully determine the exterior spacetime given the motion of the shell, exhibiting a mechanism similar to Birkhoff's theorem in vacuum GR. Finally, imposing a distributional form of the Einstein equations at the boundary hypersurface suffices to determine the shell dynamics and ensure the consistency of the solution. For simplicity, we take the bare value of the cosmological constant to be zero, so that the only contribution to the stress tensor comes from the aether. The effective time dependent cosmological constant term gives the solution a nontrivial cosmological structure.

\subsubsection{Interior solution}\label{subsubsec:interior}

The interior solution in \mmHg\;is taken to be a homogeneous isotropic spacetime
with line element $dt^2 - a^2(t)d\Sigma^2$, where $d\Sigma^2$ is the line element on 
a maximally symmetric space, with $T = T(t)$.
The $T$ field equation then holds automatically by symmetry, 
and therefore the aether stress tensor is conserved automatically, 
so the solutions are determined by the first Friedman equation alone, 
i.e., from $G_{ab}u^au^b = T^u_{ab}u^au^b$, 
where $T^u_{ab}$ is the aether stress tensor of 
Khronometric theory, \eqref{TTT}. 
The aether energy density is thus 
\begin{equation}
    \rho = - \frac{c_{\theta}}{6}\theta^2 = - \frac{3c_{\theta}}{2}\left(\frac{\dot a}{a}\right)^2\,,
\end{equation}
hence the Friedman equation becomes
\begin{equation}
    (1 + c_{\theta}/2)\dot a^2 = - k\,,
\end{equation}
where $k = 0,\pm1$ determines the type of curvature of the maximally symmetric 
spatial slices. As long as $c_{\theta}> -2$, this equation requires $k = -1$, i.e., 
hyperbolic spatial sections. 
The spacetime line element is 
\begin{equation}\label{intds2}
    ds^2 = dt^2 - \frac{t^2}{1+c_\theta/2} d\Sigma_{H^3}^2,
\end{equation}
where $d\Sigma_{H^3}^2$ is the line element on the unit hyperboloid. 

Although they look quite similar, 
there is an important difference between the metrics \eqref{intds2}  with and without nonzero $c_\theta$.
The case $c_\theta=0$ is nothing other than the Milne universe, i.e., the interior of the forward light cone of a point $p$ in Minkowski spacetime,
foliated by surfaces of constant proper time $t$ from $p$. 
The scale factor vanishes at 
the surface $t=0$, which  coincides with the light cone.
In the case $c_\theta\ne0$, in contrast, the $t=0$ surface is {\it singular}. Despite this difference, however,
all cases are conformal to the Milne universe: 
 Let $t =:\t_0(\t/\t_0)^\l$ define a new coordinate $\t$, 
where $\t_0$ is any fixed time and $\l :=1/\sqrt{1+c_\theta/2}$. 
Then $dt^2 - \l^2 t^2 d\Sigma_{H^3}^2 = \l^2(\t/\t_0)^{2(\l-1)}(d\t^2 - \t^2 d\Sigma_{H^3}^2)$.
The nonzero $c_\theta$ case is thus
a cosmological 
spacetime with an initial
singularity on a 
light cone (see Fig.~\ref{fig:sketch}).

The mean curvature of a time slice of \eqref{intds2} is $K =-\theta= -3/t$, so the aether energy density 
and pressure both diverge as $1/t^2$ as $t\rightarrow0$. 
The spacetime
curvature is proportional to $1/t^2$, so the spacetime is asymptotically flat in the future. 
The singularity at $t=0$ presents no
problem for the solution we wish to investigate, which is that of a collapsing shell of matter
whose interior is this deformed Milne universe, and which has a shell that emerges from the big bang singularity and collapses, forming a black hole. Outside the shell the universe is no longer the deformed Milne one, and we seek to solve for it.
In particular, we aim to determine whether the evolution outside the 
shell may continue after the CMC slice on which the shell collapses to a point and, if so, whether that evolution is unique. If it cannot
be continued uniquely the theory is dynamically incomplete. We will not be concerned, on the other hand, with the initial conditions, but rather take the viewpoint that the shell 
models a typical overdensity that could have formed in the early universe by some process or another. In fact, we do not even know whether, in the solutions we examine, the singular past boundary of the spacetime outside the shell remains null as in the 
interior of the shell, or if it instead becomes spacelike.

\subsubsection{Exterior Solution}\label{sec:exterior}
The exterior solution is more complicated. Our strategy is to parametrize the spherically symmetric data that satisfy the constraints for any given mean curvature  $K$, and then show that by choosing the lapse and shift appropriately these are stacked into a foliation of a spacetime that satisfies
 the evolution equations including
the matching conditions at the shell. The possible spherically symmetric CMC data for GR with a cosmological constant have been classified in \cite{cmc}, 
and this classification applies here
since the constraints are modified in \mHg\;only by the contribution of 
the (time dependent) effective cosmological constant,
\begin{equation}\label{cc}
\Tilde{\Lambda}=-\frac{c_\theta}{6}K^2\, .
\end{equation}

We adopt new coordinates in a patch around the shell given by a foliation time $t$ that agrees with the time coordinate $t$ in the interior solution \eqref{intds2}, the proper radial distance $l$ away from the shell along the foliation (chosen to be positive in the exterior and negative in the interior), and spherical angles. Since these are physically meaningful coordinates, it is important that all components of the metric tensor are continuous.\footnote{When gluing two spacetimes together along a boundary, the usual junction condition is that the \textit{pullback} of the metric agrees under the identification of the boundaries. This is natural because it is possible to vary the other components of the metric freely by performing a coordinate transformation that doesn't move any boundary points. However, in the glued manifold we still view it as physically necessary that the metric is continuous as a tensor field, meaning that its contraction with any continuous vector field is a continuous function. This amounts to requiring that its components be continuous in any \textit{differentiable} chart. The foliation-adapted $t$ and $l$ coordinates we use in the text here are intrinsically defined in such a way that we take them to constitute such a chart, which is equivalent to saying that the foliation has no kinks.} 
They are not continuously differentiable, however, due to the presence of the surface stress tensor which will produce delta functions in second derivatives of the metric that must arise from discontinuities in first derivatives.
The spacetime metric tensor  takes the form
\begin{align}\label{exterior metric}
g_{\mu\nu}dx^\mu dx^\nu
&=N^2 dt^2-(dl + Xdt)^2-r^2d\Omega^2\nonumber\\
&= (N^2-X^2)dt^2-X(dtdl+dldt)-dl^2-r^2d\Omega^2
\end{align}
with the lapse $N$, radial shift $X$, and areal radius $r$ being functions of $l$ and $t$ that are continuous everywhere.
The positive signature spatial metric tensor  and extrinsic curvature of the slices are then given by
\begin{align}\label{def uv}
   h_{ij}dx^i dx^j=dl^2+r^2d\Omega^2 
   \qquad\mbox{and}\qquad
   K_{ij}dx^i dx^j &= w\, dl^2+v\,r^2 d\Omega^2\,,
\end{align}
where $w$ and $v$ are also functions of $l$ and $t$, restricted by the CMC condition $w+2v=K$.

To solve the constraint equations, we follow the spirit of the integral form of Gauss' Law, employing quasilocal quantities on the surfaces of spheres that capture information about the stress energy distribution in the region they bound. The first of these is the Misner-Sharp energy~\cite{Misner:1964je} $E$ defined by
\begin{equation}\label{E}
1-\frac{2E}{r}=r'^2-v^2r^2
\end{equation}
This definition is convenient because it allows one to derive the following well-known reformulation of an Einstein equation \cite{Misner-Sharp} from the constraints: 
\begin{equation}\label{E'}
    E'=\frac{1}{2} r^2[T_{uu}\,r'+T_{ul}\,vr]\,,
\end{equation}
 where the indices $u,l$ represent contraction with the aether $u$ and outward pointing 
radial unit vector $\partial_l$, respectively, and 
$'$ represents directional derivative along $\partial_l$.
There is an analogous equation for the derivative of $E$ along $u$, which will be used in \ref{sec:extension}:
 \begin{equation}\label{Edot}
\dot{E}=\frac{1}{2} r^2 [T_{ul}\,r'+T_{ll}\,vr]\,,
\end{equation}
where $\cdot$ represents
directional derivative along $u$.

We can introduce a similar quantity $P$ defined on each sphere as:
\begin{equation}\label{P}
    P=\int_{S^2} K^{j}_{i}(\partial_l)^i (dl)_j dA= 4\pi r^2 w\,.
\end{equation}
This is the area flux of the vector
$K_i^j (\partial_l)^i$ through the sphere, so we can apply the covariant divergence theorem to obtain $$P(l_1)-P(l_0)=\int_{S^2\times [0,1]} D_j (K_i^j (\partial_l)^i) dV= \int 4\pi r^2[D_j K^j_i(\partial_l)^i+K^{ij}D_j(dl)_{i}] dl\,, $$
hence $P'$ is equal to the integrand of the last expression.
A straightforward computation yields that $K^{ij}D_j(dl)_{i}={v(r^2)'}/{r^2}$. Now the momentum constraints take the form $$ D_j K_i^{j}-D_iK=-T_{ui}$$ which in the case of CMC gauge simplifies to $$D_j K_i^{j}=- T_{ui}\,.$$
Thus we have the equation for the radial derivative of $P$:
\begin{equation}\label{P'}
     P'=4\pi r^2 \left [- T_{ul}+\frac{v(r^2)'}{r^2}\right ]\,.
\end{equation}
This is equivalent to the radial momentum constraint, and the angular components are satisfied identically in spherical symmetry.

In our case $T_{ul}$ vanishes, and $T_{uu}=\Tilde{\Lambda}$. 
Equality of the derivative of \eqref{P} with \eqref{P'} implies that $(wr^2 )' = v(r^2)'$. 
Since $w+2v=K$ is constant on a slice 
we have $w=\frac13 K + 2\alpha$ and $v=\frac13 {K}-\alpha$
for some function $\alpha$, which satisfies $\alpha'= -{3\alpha} r'/{r}$. It follows that
\begin{equation}\label{C}
   w=\frac{K}{3}+\frac{2C}{r^3},\;\;\;v=\frac{K}{3}-\frac{C}{r^3}
\end{equation}
where $C$ is an integration constant. Similarly, the equation for $E'$ (which amounts to the Hamiltonian constraint in this case) yields 
\begin{equation}\label{M}
    r'^2=1-\frac{2M}{r}-\frac{\Tilde{\Lambda}}{3}{r^2}+v^2r^2
\end{equation}
with an integration constant $M$ (which is related to to the Misner-Sharp energy by $E=M+{\tilde{\Lambda}}r^3/6$). 
These equations agree with those in \cite{cmc}, where it is also  
shown that the relevant solutions to \eqref{M}  are typically periodic in $l$. In this case, one can consider the periodic solution, or instead glue an identical collapsing shell with an interior region at another "end". This will not affect the subsequent discussion, which is concerned with the time evolution of the parameters describing the data as well as the location of the adjoining boundary. The parameters $M$ and $C$ at each time will be determined in the following subsection by requiring that the exterior is consistently glued to the interior at the shell while satisfying the junction conditions.
\par

For now, suppose that $M$ and $C$ are simply prescribed at each time. Does there exist a unique exterior spacetime (not a priori satisfying the evolution equations) such that on each time slice the data is given by the preceding solutions to the constraint equations, and with these boundary conditions? The answer is yes, subject to one consistency condition. Note that constructing such a spacetime in these
coordinates amounts to solving for the lapse $N$ and shift $X$ 
such that the specified tensor $K_{ij}$ is in fact the extrinsic curvature --- the other functions 
are specified throughout the exterior region via the solution of the constraint equations. This amounts to two equations,
\begin{equation}\label{lapse+shift}
    w = X'/N \;\;\;\;\;\;         v = -\dot r/r = (1/Nr)(-r_{,t} +Xr')\,,
\end{equation}
which are differential equations that relate the metric components to the 
given extrinsic curvature functions $w$ and $v$.
The second equation algebraically relates $N$ and $X$, and the first 
provides a first order ODE which can then be solved uniquely for either $N$ or $X$ given a boundary value. 
However, the interior solution fixes boundary values for \textit{both} the lapse and shift, which must be consistent
with the second equation. That equation 
holds automatically in the interior, since
that is a prescribed spacetime, and 
the partial derivative $r_{,t}$ 
is continuous, because the shell
has a well-defined intrinsic geometry, 
hence there must be no jump in the 
quantity $Xr'-Nrv$. 
The discontinuities 
across 
the boundary, $\Delta v$ and $\Delta r'$, 
must therefore satisfy 
\begin{equation}\label{consistency1}
    X\Delta r'=NR\,\Delta v\,.
\end{equation}

Now we turn to the conditions that must be satisfied for the constructed exterior spacetime to be a solution of the Einstein equations with the aether stress tensor. In spherical symmetry the
independent components of the field equations correspond, in our coordinates,
to $E_{tt} = E_{tl} = E_{ll} = E_{\theta\theta}=0$, in the notation 
of \eqref{T-diffeo2}.
 The first two are the constraints, which are satisfied by the above construction. The latter two are equations of motion which must still be imposed.  
Since $K = -\theta$ is constant on the preferred slices, the $T$ field 
satisfies its field equation \eqref{mTeqn},
and thus $\nabla_a E^{ab}=0$, even if the constructed spacetime is off-shell.
This provides two independent identities, 
$\nabla_\a E^{\a l}=0$ and 
$\nabla_\a E^{\a t}=0$. Written out explicitly, 
and imposing the constraints $E^{tt}=E^{tl}=0$ and spherical symmetry, these 
take the form 
\begin{align}
\Gamma^t{}_{ll}E^{ll} + 2\Gamma^t{}_{\theta\theta}E^{\theta\theta}&=0 \\
\partial_l E^{ll} + \Gamma^\a{}_{\a l}E^{ll}  + \Gamma^l{}_{\theta\theta}E^{\theta\theta}&=0
\end{align}
The $t$-identity allows one to solve for $E_{\theta\theta}$ in terms of $E_{ll}$ and the Christoffel symbol.\footnote{This argument can fail when $\Gamma^t_{\theta\theta}=0$. However this Christoffel symbol is equal to $-r{\dot{r}}/{N}={vr^2}/{N}$, which does not vanish except possibly at isolated points. Therefore in our case this issue can be safely ignored.}. 
The $l$-identity is then 
an ODE for $E^{ll}$ with coefficents constructed out of Christoffel symbols, which then
implies that if $E^{ll}$ vanishes
at the exterior limit of the shell, 
it and $E^{\theta\theta}$ vanish throughout the exterior. This guarantees that
\textit{all of the Einstein equations hold throughout the exterior}! Therefore just one component of the dynamical equations needs to be imposed at the exterior limit in order for the constructed stack of CMC slices to be a valid exterior solution to \mmHg. For instance, we could impose equation \eqref{Edot}. However, we will hold off on doing this, because it will turn out that this remaining component can be incorporated into the equations describing the evolution of the spherical shell.

\subsubsection{Junction Conditions}\label{J C}
Finally, it is necessary to impose the Einstein equations distributionally \textit{at} the matter shell, not just in the exterior and interior separately. This is equivalent to imposing the secondary Israel-Darmois junction conditions. In spherical symmetry there are four independent components to the Einstein tensor, so four equations must be imposed relating the surface stress tensor to delta function singularities (or the absence thereof), in addition to forcing the step function discontinuities to vanish (which yields one equation as discussed above). 

The matter is taken to be a spherical dust shell with radial 4-velocity $\chi$,
with  stress tensor 
\begin{equation}\label{shellT}
   T_{ab}=-\sigma {\chi_a \chi_b}\delta(l)/(\partial_l\cdot n) 
\end{equation}
where $\sigma$ is the dust rest mass area density 
and $n$ is the outward-pointing unit normal to the timelike shell worldtube.

Expanding $\chi$ and $n$ in the orthonormal basis $\{u, \partial_l\}$ adapted to the preferred foliation one has
\begin{align}
\chi &= \chi^u\, u + \chi^l\, \partial_l = (Nu + X \partial_l)/\sqrt{N^2 - X^2}\label{chiu}\\
n &= \chi^l\, u + \chi^u\, \partial_l\,.
\end{align}
The second equality in \eqref{chiu} 
holds since the radial length coordinate $l$ has been set to zero at the shell, 
so $\chi$ is the unit vector parallel to $\partial_t= Nu + X \partial_l$.
The projected stress tensor components that enter the constraints are
given by
\begin{equation}
    T_{uu}:= T_{ab}u^a u^b = \sigma \chi^u\delta(l) \;\;\;\;\;\;\;
    T_{ul}:= T_{ab}u^a \partial_l^b= -\sigma \chi^l\delta(l)
\end{equation}
which must be matched to delta function singularities in the 
corresponding Einstein tensor components. 
The ${uu}$ and ${ul}$ components of the Einstein equation 
 are the constraints; they set the jumps in $E$ and $P$ across the boundary, 
 and therefore also the jumps in $r'$, $w$, and $v$. 
 This mathematical problem is studied in the Appendix, where
 these jumps are found to be
 \begin{equation}\label{jumps}
 \Delta r'=-\frac{1}{2}R\sigma\chi^u\;\;\;\;\;\;\;\;\;\;\;\;\;\;\;\;\;\;\;\;\;
   \Delta w = \sigma \chi^l
\;\;\;\;\;\;\;\;\;\;\;\;\;\;\;\;\;\;\;\;\;
     \Delta v = -\frac{1}{2}\sigma \chi^l
 \end{equation}
 Thanks to the form of the surface stress tensor, these identically satisfy equation \eqref{consistency1}\footnote{This is not a coincidence. The consistency condition \eqref{consistency1} arose from imposing the continuity of the metric across the boundary. In \cite{MTW} (see page 553) it is argued that the continuity of the metric implies that the surface stress tensor cannot have a flux normal to the boundary, meaning that the matter constituting the shell must be confined to the boundary. Formally, $n\cdot T=0$. This is ensured for the dust stress tensor under consideration because $\chi\cdot n=0$.}. 
 
 Three equations remain to be imposed: the distributional 
 $G_{ll}$ and $G_{\theta\theta}$ equations on the boundary, 
 and the aforementioned single component of the Einstein equation at the exterior shell boundary. 
 These determine the motion and density of the shell throughout its trajectory, and are universal for thin dust shells. They can be reformulated in the general formalism found, e.g., in section 21.13 of \textit{Gravitation} by Misner, Thorne, and Wheeler (see exercise 21.26) \cite{MTW}. According to our conventions, these are
\begin{equation}\label{Shell EOM}
4\pi R^2 \sigma =\mu\;\;\;\;\;\;\;\;\;\;\;\;
\a_+ + \a_-=0\;\;\;\;\;\;\;\;\;\;\;\;
\a_+ - \a_-=\frac{1}{2}\sigma n
\end{equation}
where $\mu$ is a constant which gives the "total rest mass" of the shell, and $\a_-,\a_+$ are the 4-accelerations $\nabla_\chi \chi$ as computed in the inner and outer geometries respectively. The latter two may appear to be vector equations, but since $\a_+,\a_-$ are multiples of $n$ in spherical symmetry, they each have only one relevant component.

\subsubsection{Collapse Evolution Equations}
The 
equations \eqref{Shell EOM}, together with the known interior solution, can be used to derive a pair of coupled first order ordinary 
differential equations controlling the motion of the shell. 
First note that 
\eqref{Shell EOM} implies $\a_-\cdot n =\sigma/4={\mu}/{16\pi R^2}$. 
Another expression for $\a_-\cdot n$ can be computed
directly, using the facts that $N=1$ and $u$ is geodesic 
in the interior geometry \eqref{intds2}.
This computation results in
$\a_-\cdot  n=-\partial_t{\chi}^l+w_-\chi^l$.
Equating the two expressions for 
$\a_-\cdot n$ yields one ODE, 
$\partial_t{\chi}^l= w_-\chi^l - {\mu}/{16\pi R^2}$. 
A second ODE is obtained from  \eqref{lapse+shift}, which evaluated at the interior side of the shell yields $\partial_t {R}=(r'{X}-v_-r)|_R$.
Using the hyperbolic space metric on the interior we find 
$r'|_R=\sqrt{1+{R^2}/{a^2}}$,
using $\chi = \partial_t/\sqrt{1-X^2}$
we find $X|_R = \chi^l/\sqrt{1 + (\chi^l)^2}$, and since the interior is regular at $r=0$
 the parameter $C$ in \eqref{C} vanishes so we have $w_-=v_-={K}/{3}$. The two coupled first order ODEs 
 describing the motion thus take the form
\begin{equation}\label{Motion Equations}
    \partial_t{\chi}^l=\frac{K}{3}\chi^l-\frac{\mu}{16\pi R^2},\qquad \qquad\partial_t{R}=\pm\sqrt{\frac{1+(R/a)^2}{1+(\chi^l)^{-2}}}-\frac{1}{3}KR\,,
\end{equation}
where $\pm$ corresponds to the sign of $\chi^l$, and from \eqref{intds2} we have the explicit time dependence 
$K(t)=-{3}/{t}$ and 
$a(t)=t/\sqrt{1 + c_\theta/2}$.

The coupled evolution equations are complicated, but given initial conditions they
can be solved  numerically 
to determine the motion of the shell. After this the exterior spacetime can be constructed slicewise by using equations \eqref{jumps} to extract the parameters $M(t)$ and $C(t)$, both of which  vanish inside the shell, and numerically integrating  equation \eqref{lapse+shift} outward from the shell on each timeslice, to find the lapse and shift. 
Equation \eqref{C} 
together with \eqref{jumps} and 
\eqref{Shell EOM}
yields 
\begin{equation}\label{C_solution}
    C(t) = 
    \frac12 R^3 \Delta w =\frac{\mu}{8\pi}R\chi^l
\end{equation}
To find $M(t)$, we start 
by solving
\eqref{M} for 
$M = \frac12 r(-r'^2 + v^2 r^2 + 1 - \tfrac13 \tilde\Lambda r^3)$. 
Since $M$ vanishes inside the shell, 
and there is no jump in 
$1-\tilde \Lambda r^2/3$, we have
\begin{align}
    M(t) &= \frac12 R\Bigl(R^2\Delta(v^2)-\Delta(r')^2\Bigr)\label{M2a}\\
    &=\frac12 R\Bigl(R^2(\Delta v)^2 + 2R^2v_-\Delta v-(\Delta r')^2 - 2r'_-\Delta r'\Bigr)\label{M2b}\\
    &= R\Bigl(-\frac18\sigma^2 R^2+ R^2v_-\Delta v -r'_-\Delta r'\Bigr)\label{M2c}\\
      &= \frac{\mu}{8\pi} \Bigl(-\frac{\mu}{16\pi}R^{-1}+r'_-\chi^u - v_- \chi^lR\Bigr)\,,\label{M2d}
\end{align} 
where we used the jump relations \eqref{jumps}
to combine the first and third terms
in \eqref{M2b}
using the normalization of $\chi$, and for the substitutions leading to \eqref{M2d}. 
The coefficients $v_-$ and $r'_-$ are 
obtained using
\eqref{C}, \eqref{M}, and \eqref{cc}:
\begin{align}
    v_- &= \frac{K}{3}\\
    r'_-&= \sqrt{1 + \left(\frac19 K^2 -\frac13\tilde\Lambda\right) R^2} 
    = \sqrt{1 + \frac19 (1 + c_\theta/2)K^2 R^2}  
\end{align}

\subsubsection{Late Time Behaviour}
We will now focus on the behaviour late in the collapse process,
where the evolution equations can be simplified.
Suppose that $R$ is small compared to relevant cosmological length scales and $\chi^l$ is large and negative, so the shell is contracting inwards, so that
\begin{equation}\label{approximations}
    R\ll a, \qquad R\ll 1/{|K|},\qquad \chi^l\ll -1\,.
\end{equation}
The $\partial_t R$ evolution equation in \eqref{Motion Equations}
then takes the approximate form
$\partial_t R=-1$, which 
yields the solution 
$R\approx t_f-t$ (where $t_f$ is the time when the shell collapses to a point). 
The $1/R^2$ term in the $\partial_t\chi^l$ equation
must overpower the $\chi^l$ term, since 
otherwise $O(\chi^l)\ge O((t_f-t)^{-2})$,
which would imply 
$O(\partial_t\chi^l)\ge O((t_f-t)^{-3})$,
which would be inconsistent with 
\eqref{Motion Equations}.
The velocity component $\chi^l$ is thus driven to $-\infty$ by the $1/R^2$ term, 
and the asymptotic solution takes the form 
\begin{equation}\label{Late time collapse solution}
R(t)=t_f-t,\qquad \chi^l(t)=-\frac{\mu}{16\pi}\frac{1}{t_f-t}+\mbox{const.}
\end{equation}
This solution is applicable in regimes where equations \eqref{approximations} are valid. 
 
As the shell radius goes to zero, the quantities $C(t)$ and $M(t)$ remain finite. The limit of $C(t)$ is given by plugging \eqref{Late time collapse solution} into \eqref{C_solution}:
\begin{equation}\label{Clim}
    C(t) \overset{t\rightarrow t_f}{\longrightarrow}\; -\frac12\left(\frac{\mu}{8\pi}\right)^2
\end{equation}
The asymptotic solution for $\chi^l$ 
given in \eqref{Late time collapse solution}
shows that $\chi^u\rightarrow \frac{\mu}{16\pi}R^{-1} + \mbox{finite}$, so 
the third term in \eqref{M2d} remains finite and the 
divergences in the first and second terms cancel, 
thus $M$ remains finite. To evaluate this
finite value we need to keep track of subleading terms. 

With $R$ as the expansion parameter, we've seen that the $R^{-1}$ terms in equation \eqref{M2d} for $M(t)$ cancel. 
The $R^0$ terms come only from two places: 
the $R^0$ term in $r'_-$ times the $R^0$ term in $\chi^u$, and the $O(R^{-1})$ term in $\chi^l$.
Thus we have
\begin{align}
    M \longrightarrow 
    \frac{\mu}{8\pi} \Bigl(\chi^u_0 - \frac{\mu}{16\pi t_f}\Bigr)
    \label{Mlim}
\end{align} 
The integration constant $\chi^u_0$ is determined implicitly by the initial data. If it is not sufficiently large, the limiting $M$ value would be negative. Since $M$ is the limit of the Misner-Sharp energy $E$ (see equation \eqref{Elim_M}) approaching the center, and the condition for a trapped surface is $E>\frac 12 R$ \cite{Misner-Sharp}, spheres close to the singularity are not trapped in this case.
It could be interesting to explore the
relation between initial conditions for the collapse and the limiting
sign of $M$, but we have not done so.

In summation, the collapsing solution ends at a final slice, where $M$, $C$, and $K$ take finite values. The spatial metric and extrinsic curvature 
both have a well-defined limit on this slice, which can be obtained by substituting \eqref{Clim} and \eqref{Mlim} into equations \eqref{C} and \eqref{M}, 
which can be solved for the function $r(l)$ in the metric \eqref{exterior metric}.
There is a curvature singularity at the center, but the  spacetime is otherwise smooth at the final slice. This constitutes the end of the regime where the spacetime is completely determined by the collapse equations \eqref{Motion Equations}.
 
\subsubsection{Extending After Collapse}\label{sec:extension}
There remains the question of whether the solution can be extended past this final time, using the data on the final slice as initial conditions. Due to the infinite propagation speed of the elliptic mode, this slice appears to act as a kind of Cauchy horizon; information can in principle travel out of the singularity to affect the slices to its future, even if they are spacelike separated from it. Therefore the dynamics seems ill-posed unless further assumptions are made. We will suppose that the metric and preferred foliation remain spherically symmetric, with slices extending to the singularity at the center $r=0$,  and we define the coordinate $l$ to be proper radial distance from $r=0$ on those slices (which according to \eqref{M} is finite). The analysis of the constraint equations in section \ref{sec:exterior} still applies, so the data on each slice are parametrized by $M$ and $C$, with $K$ a specified function of the time coordinate $t$ (which is no longer the interior proper time of section \ref{subsubsec:interior}). A priori $K$ can now be either in the monotonic or the constant regime as discussed in \ref{sec:cauchy}. The lapse and shift also obey equations \eqref{lapse+shift}, which we can use to conduct an asymptotic analysis of their behaviour near the singularity. 
Using equations \eqref{C}, \eqref{M}, and \eqref{lapse+shift}, retaining only leading terms in $1/r$ we find
\begin{equation}
    -\frac{NC}{r^2}\sim\frac{X|C|}{r^2},
\end{equation}
since $r_{,t}$ goes to zero approaching the center and $C$ is nonzero in both equations \eqref{C} and \eqref{M}. As $C<0$ near the slice on which the 
shell collapses to a point (see \eqref{Clim}), this implies that in the limit as $r\rightarrow 0$, ${X}/{N}\rightarrow 1$. Now considering the equation for $w$, at leading order it reads
\begin{equation}
    X'\sim\frac{2NC}{r^3},
\end{equation}
or expressed in terms of a derivative with respect to $r$,
\begin{equation}
    \frac{dX}{dr}\sim\frac{2NC}{r|C|}.
\end{equation}
Integrating this for $X$, we find that the equation is consistent with the previous conclusion that $X\sim N$ only if $N,X\sim r^{-2}$ as $r\rightarrow 0$.
\par
Although there is no longer a shell, and so there are no junction conditions to impose, the argument from the Bianchi identities at the end of section \ref{sec:exterior} still holds and the full Einstein equations follow from imposing,
besides the constraints,
one additional component of the Einstein equations at one regular spatial point on each slice.
We will extract this equation from the relation \eqref{Edot} for the proper time derivative of the Misner-Sharp mass, which is an evolution equation rather than a constraint. From \eqref{TTT2} we have $T_{ul}=0$, $T_{uu}= -\frac{c_\theta}{6}K^2$,  and $T_{ll}=\frac{c_\theta}{3}(\frac{1}{2}K^2-\dot{K})$. Substituting these and the expression \eqref{C} for $v$ into \eqref{Edot}, we find 
\begin{equation}\label{Edot2}
\dot{E}=\frac{c_\theta}{6}(\frac{1}{2}K^2-\dot{K})(\frac{K}{3}-\frac{C}{r^3})r^3.
\end{equation}
Note that equations \eqref{E} and \eqref{M} imply
\begin{equation}\label{Elim_M}
    E-\frac{\tilde{\Lambda}}{6}r^3 = M.
\end{equation}
Differentiating \eqref{Elim_M} with respect to proper time yields (using \eqref{lapse+shift})
\begin{equation}\label{Edot3}
    \dot{E}-\frac{\dot{\tilde{\Lambda}}}{6}r^3+\frac{\tilde{\Lambda}}{2}vr^3 = \dot{M}.
\end{equation}
This equation can be combined with \eqref{Edot2} to get
\begin{equation}
   \dot M = \frac{c_\theta}{6}C\dot K
\end{equation}
at any point, which then implies
\begin{equation}
    \partial_t M = \frac{c_\theta}{6}C\,\partial_t K,
\end{equation}
since $M$ and $K$ are constant on a
constant $t$ surface.
This fulfills the role of the single component equation and guarantees that the full Einstein equations hold everywhere. 
\par
The evolution of $M$ 
is thus determined after prescribing $K$ and $C$ as arbitrary functions of time. Assuming that we remain in the regime where $K$ is monotonic, it is clear that this evolution is nonunique, which means that the Cauchy problem is ill-posed beyond the point of collapse \textit{even with the assumption of spherical symmetry}. This affects not only the determination of the foliation, but the spacetime geometry itself, which is apparent from the arbitrariness of the mass parameter that describes the central limit of the Misner-Sharp energy.

\section{Conclusion}

We have considered the dynamics of the IR limit of 
Ho\v{r}ava gravity in the case where
of the three coupling parameters in the theory 
only the expansion coupling is nonzero. 
For general boundary conditions the dynamics
of this theory contains a half-mode with 
pathological dynamics which is eliminated by the
requirement that the preferred foliation consist of 
CMC surfaces. The resulting theory, which we call minimal minimal 
Ho\v{r}ava gravity,
$\mathrm{m^2Hg}$, 
{is equivalent to the Cuscuton model with quadratic potential for the scalar
Cuscuton field \cite{Afshordi:2006ad, Afshordi:2009tt,
Bhattacharyya:2016mah}, and to 
Einstein-aether theory with all aether couplings except $c_\theta$ set to zero,
restricted to the sector in which the aether expansion is nonzero
and the surfaces of constant expansion are spacelike \cite{Bhattacharyya:2016mah}.}

\par
The resulting initial value problem consists of a CMC evolution with "time-varying cosmological constant" $\tilde{\Lambda}=-\frac{c_\theta}{6} K^2$ in the Hamiltonian constraint. The dynamics can be classified into two cases, where $K$ is either time-independent or monotonic respectively. In the former case the theory reduces to GR with cosmological constant equal to $\tilde{\Lambda}$ but restricted to spacetimes admitting a CMC foliation. In the latter case 
the aether stress tensor has the form of a perfect fluid
so the theory differs from GR. Focusing on this case, we constructed spherical collapse solutions by gluing an interior FLRW solution to an exterior spherically symmetric spacetime across a gravitating thin shell of dust. After selecting a particular interior geometry, we found that the exterior geometry is completely determined by the motion of the shell, analogous to Birkhoff's theorem. Interestingly, this result is obtained by applying conservation identities that apply off-shell (meaning when the constraints but not necessarily dynamical equations are satisfied). The remaining equations of motion reduce to a system of coupled ODE's describing the motion of the shell. 
\par
Analyzing these equations in the late time limit, we found that the shell collapses to a point on a particular slice of the preferred foliation. The intrinsic and extrinsic geometry of this slice are completely determined by three parameters; the mean extrinsic curvature $K$, the Misner-Sharp energy of the collapsed shell $M$, and a third parameter $C$  that in this limit depends only on the "rest mass" of the shell. The solution can be extended beyond this point while obeying the $\mathrm{m^2Hg}$ field equations, but this extension is not unique. As long as $K$ continues to be monotonic, the evolution of $M$ can be specified arbitrarily, which then determines $C$. One can contrast this situation with the indeterminacy of perturbations to the Reissner-Nordstrom black hole. Although that solution of GR has a Cauchy horizon because information can "come out of the singularity", if one restricts to spherical symmetry the only allowed solution is the Reissner-Nordstrom solution itself, with prescribed mass and charge. In the case of $\mathrm{m^2Hg}$, in the presence of a singular center there is a single spherical degree of freedom with undetermined time evolution, which is absent in Einstein-Maxwell theory. This provides evidence that $\mathrm{m^2Hg}$ is not a physically viable theory, as the black holes we have observed in nature do not seem to present such behaviour (for instance, the pattern of gravitational waves emitted from black hole mergers are in accordance with the deterministic predictions of GR).
\par
Some questions remain to be answered. We have not examined in detail the signature of the singularity that forms. Due to the infinite propagation speed, the singularity can give rise to an indeterminate evolution without necessarily being timelike.  The fact that $N/X\rightarrow 1$ as $r\rightarrow 0$ \textit{suggests} that it may be null,  but since the metric is 
singular a more careful analysis is required to establish this.
Moreover, we do not know the asymptotic structure of the solution, even prior to the collapse. In the far past when the shell is dispersed, it likely resembles that of the interior geometry, which is conformally equivalent to the Milne universe. However when the shell density is sufficient to induce significant backreaction, the asymptotic structure can only be determined after solving for the metric. This can be done for any solution to the collapse equations \eqref{Motion Equations} by integrating the ODE \eqref{M} slicewise to determine the spatial metric, and then either integrating \eqref{lapse+shift} for the lapse and shift \textit{or} directly solving \eqref{lapse} (which reduces to an ODE in spherical symmetry) for the lapse. 
Knowledge of the asymptotic structure in
the vicinity of the slice of collapse 
would allow one to
address the question of whether a
\textit{universal horizon}
forms, that is, whether 
slices of the Ho\v{r}ava evolution,
when parametrized by an infinite proper time range of asymptotic 
observers, pile up in the center of the collapsing shell 
and stop before a certain regular final slice \cite{SergeyPeyresq,Barausse:2011pu,
Blas:2011ni,Bhattacharyya:2015gwa}.
If that were to happen, the non-unique evolution
entailed by the singularity would at least be hidden to 
asymptotic observers. 

We also do not know if
spherical collapse solutions will continue to
exhibit nonunique evolution if small but nonzero
values of the shear and/or acceleration terms in the
khronon Lagrangian, or the higher derivative
terms in the complete Ho\v{r}ava theory are included.
Since the latter terms are essential to the
UV regulating motivation for the Ho\v{r}ava proposal, 
it would be important to understand their consequences
for predictivity in order to
establish whether the proposal is indeed
not viable.

\par
\section*{Acknowledgments}
We thank Niayesh Afshordi, Rodrigo Andrade e Silva, Enrico Barausse, 
Noemi Frusciante, Sean Gryb, Jim Isenberg, David
Mattingly,
Shinji Mukohyama, Sergei Sibiryakov, Thomas Sotiriou, and Chulmoon Yoo
for useful discussions.

This research was supported in part by 
NSF grants PHY2012139 and
PHY2309634, and by Perimeter Institute
where part of this work was carried out. Research at Perimeter Institute is supported in part by the Government of Canada through the Department of Innovation, Science and Economic Development
and by the Province of Ontario through the Ministry of Colleges and Universities.

\appendix
 
\section{Jump Equations}
In this appendix we derive the jump equations \eqref{jumps}. 
The starting point is the stress tensor for a pressureless dust shell (see, e.g., \cite{poisson}): 
\begin{equation}\label{shellT1}
    T_{ab}=\sigma u_a u_b\delta(\tilde{n}), 
\end{equation}
where $\tilde{n}$ is the proper length parameter along any congruence passing perpendicularly through the shell's $D$ dimensional hypersurface (in a $D+1$ dimensional spacetime). The tangent vectors to the congruence at each point on the surface are the outward pointing unit normals $n$, which satisfy $n\cdot d\tilde n = 1$. We would like to convert the variable in the delta function to the coordinate $l$ that measures proper length on the CMC slices. 
To do this, we first spell out explicitly a definition
of such a delta function,
and then apply it to the present case.

To integrate any top form 
$\omega$ against $\delta(f)$ means to integrate the $(D-1)$-form $v_f\cdot\omega$ 
over the $f=0$ hypersurface $\Sigma$, 
where $v_f$ is any vector satisfying $v_f\cdot df =1$. This is independent
of the choice of $v_f$, because if $v_f + s$ is another such vector then 
$s$ is tangent to $\Sigma$ so the pullback of $s\cdot \omega$ to the 
$\Sigma$ vanishes. Now if $g$ is another function 
whose $g=0$ hypersurface agrees with that of $f$, 
then $\delta(f) =\delta(g)/(v_g\cdot df)$,
since $v_f = v_g/(v_g\cdot df)$ (modulo vectors tangent to $\Sigma$).
In the present case,
since $\tilde n$ and $l$ have the same zero hypersurface,
it follows that 
\[
\delta(\tilde n)=\delta(l)/(\partial_l\cdot d\tilde n) =-\delta(l)/(\partial_l\cdot n),
\]
(where the final $\cdot$ denotes the metric inner product), and hence 
the stress tensor \eqref{shellT1}
can be expressed as  
\begin{equation}
   T_{ab}=-\sigma {\chi_a \chi_b}\delta(l)/(\partial_l\cdot n)\, .
\end{equation}
which is the form 
introduced in \eqref{shellT}.

The delta functions in the stress tensor components
\textit{cannot} be plugged directly into the equation
\eqref{E'}
for the radial derivative of $E$, since that equation assumes more regularity of the quantities involved. (Indeed, $T_{uu}r'$ would involve the undefined product of
a delta function with a step function).
Instead we must return to the general expressions for the 
Einstein tensor in spherical symmetry, which can be found 
in \cite{symmetry}. 
Let $\gamma$ be the radial coordinate, equal to $l$ on an initial slice, and 
extended as constant along the normals to the CMC foliation, so that the 
shift vector vanishes in the $(t,\gamma)$ coordinate system.
The relevant components on the initial slice are: 
\begin{equation}\label{G_uu}
G_{uu}=-\left(\frac{\dot{r}^2}{r^2}+2\dot{l_\gamma}\frac{\dot{r}}{r}\right)-\frac{1}{r^2}+\left(2\frac{r''}{r}+\frac{r'^2}{r^2}\right)
\end{equation}
\begin{equation}\label{G_ul}
G_{ul}=\frac{2}{r}\left(\dot{r}'-\dot{l_\gamma}r'-\frac{N'}{N}\dot{r}\right),
\end{equation}
where differentiation with respect to $\gamma$ is denoted via subscript. 
The only term in $G_{uu}$ admitting a delta function singularity is the one involving $r''$, so integrating the $uu$ component of the Einstein equation across the boundary yields
\begin{equation}\label{r'jump}
   -\sigma \chi^u=\frac{2}{R}\Delta r'\implies \Delta r'=-\frac{1}{2}R\sigma\chi^u
\end{equation}
Similarly, the only possible delta function in $G_{ul}$ comes from $\dot{r}'$, which 
yields
\begin{equation}\label{rdot jump}
    \sigma \chi^l = \frac{2}{R}\Delta \dot{r} \implies \Delta v = -\frac{\Delta \dot{r}}{R}=-\frac{1}{2}\sigma\chi^l,
\end{equation}
using $v=-\dot{r}/r$.
Finally, from $w+2v=K= {\rm constant}$ we see that $\Delta w = -2\Delta v$. 
In summary, the jump conditions are given by
\begin{equation}\label{jumps2}
 \Delta r'=-\frac{1}{2}R\sigma\chi^u\,,\qquad
   \Delta w = \sigma \chi^l\,,\qquad
     \Delta v = -\frac{1}{2}\sigma \chi^l\,.
 \end{equation}

\bibliographystyle{JHEPmod}
\bibliography{refs}

\end{document}